\documentclass[12pt]{article}
\usepackage[utf8]{inputenc}
\usepackage{xcolor}
\usepackage{float}
\usepackage[justification=centering]{caption}
\usepackage[margin=0.5in]{geometry}
\usepackage[pdftex]{graphicx}
\graphicspath{{\Documents}}     
\DeclareGraphicsExtensions{.pdf,.jpeg,.png}
\usepackage{titlesec}
\usepackage{amssymb,amsmath}
\usepackage{moreverb}
\usepackage{multirow}
\usepackage[utf8]{inputenc}
\usepackage{natbib}
\usepackage{multicol}

\newcommand{\mcT}{\mathcal{T}}

\title{\textbf{Modern Non-Linear Function-on-Function Regression}}

\begin{multicols}{2}
\author{Aniruddha Rajendra Rao\\
Department of Statistics\\
\vspace{1cm}
Pennsylvania State University, USA\\
\and
Matthew Reimherr\\
Department of Statistics\\
Pennsylvania State University, USA}
\end{multicols}

\date{}

\begin{document}

\maketitle
\vspace{-1.5cm}
\hspace{0.35cm}\textbf{Keywords:}   Functional Data Analysis, Deep Learning, Function-on-Function Regression, Functional Neural Network, Functional Response.

\abstract{We introduce a new class of non-linear function-on-function regression models for functional data using neural networks. We propose a framework using a hidden layer consisting of continuous neurons, called a continuous hidden layer, for functional response modeling and give two model fitting strategies, {Function-on-Function} Direct Neural Networks (FFDNN) and {Function-on-Function} Basis Neural Networks (FFBNN). Both are designed explicitly to exploit the structure inherent in functional data and capture the complex relations existing between the functional predictors and the functional response. We fit these models by deriving functional gradients and implement regularization techniques for more parsimonious results. We demonstrate the power and flexibility of our proposed method in handling complex functional models through extensive simulation studies as well as real data examples.}

\section{Introduction}\label{sec1}

Functional data has gained a lot of attention in the past decades \citep{ramsay1997functional,FDA,10,hor} with advancement in data collection and storage technologies. Data that are represented as curves varying over a continuum (time, space, depth, wavelength, etc.) have been encountered widely in several areas such as Computer Vision, Time Series, Natural Language Processing, Genetics, Finance, Meteorology, Longitudinal Studies, and many more. There is a rich amount of literature in Functional Data Analysis (FDA) ranging from topics like Functional Principal Component Analysis, Canonical Correlation Analysis, Functional Regression, etc.

Functional Linear Models (FLM) are the most well-studied models in FDA. We can distinguish these models as scalar-on-function, function-on-scalar, and function-on-function regression. When discussing such models, the first term indicates the type of the response and the second term indicates the type of the predictor. In contrast to Functional linear models, which have been widely studied and investigated, there has been substantially less work done in the area of non-linear functional regression models. It is natural to begin examining non-linear models after establishing the linear methods but there is also a practical need for such models as the data may contain complicated functional relations and interactions \citep{articlel3}. Such type of data are common in practice.

In FDA, non-linear regression is an active area of research but most of the work concentrates on the scalar-on-function setting, while the function-on-function setting has picked up more interest recently. Some common ways of dealing with non-linearity in scalar-on-function setting are by using the Continuously Additive Model (CAM) \citep{McLean,articlel3,articlel4,articlel1,articlel2,reimherr2017optimal}, Quadratic Model \citep{quad}, Single Index Model \citep{s1,EILERS2009196,bookr1}, Multiple Index Model \citep{articlem1,articlem2,articlem3}, Functional Neural Network \citep{1007599, articler4,ConanGuez2002MultilayerPF,wang2019multilayer,wang2019remaining} and Functional Direct Neural Network/Functional Basis Neural Network \citep{rao2021nonlinear}. The above methods are analogous to their multivariate counterparts and most of statistical methods suffer from the drawback of either choosing to capture the complex non-linear relations or the higher order interactions. This is because they assume a very particular structure, which limits the kind of non-linear patterns that can be learned, especially in terms of interactions within the time points for the predictors. \cite{articler1} and \cite{articler1m} have discussed most of the above approaches in their review papers. The deep learning methods like Functional Neural Network and the architecture proposed in \cite{rao2021nonlinear} don't have such limitations and have been proven to work well when the true nature of the functional data is unknown.

When it comes to function-on-function regression, \cite{Lian,Kadri} extended on the work by {\cite{PREDA2007829}}, where they fit a fully non-linear model using Reproducing Kernel Hilbert Spaces (RKHS) to functional outcomes. More recent works are by \cite{Scheipl}, \cite{Kim1}, \cite{Sun}, and \cite{SUN2020106814}, where \cite{Scheipl} gave an additive model for function-on-function regression based on \cite{articlel1} and fitted it using trivariate tensor product basis approach. \cite{Kim1} introduced another estimation approach to fit the same model using orthogonal B-splines and Functional Principal Component Analysis (FPCA). The work by \cite{Sun} is an extension of \cite{Caiff} where they use an RKHS framework to estimate a function-on-function linear model. The functional quadratic model by \cite{quad} was extended to the functional response case by \cite{SUN2020106814}. Also, functional single index models \citep{s1,EILERS2009196,bookr1} were extended to the function-on-function setting by \cite{Jiang5}. All these methods help fit different function-on-function models using basis expansion or RKHS but they are still restrictive, either they are linear models or ignore interactions. They all have a particular structure that limits the kind of patterns they can learn.

From a deep learning view, \cite{1007599, articler4,ConanGuez2002MultilayerPF} first introduced the Functional Neural Network (FNN) and proved its validity by theoretical arguments and simulation experiments. FNN was then further investigated by \cite{wang2019multilayer,wang2019remaining} using FPCA. They also demonstrated the superiority of FNN to other deep learning alternatives (CNN, RNN, LSTM). This expansion of the neural network to functional data enabled efficient non-linear learning when the predictors are functions. In the FNN architecture, only the first layer is a functional hidden layer with functional neurons that give a scalar output, which is then connected to regular hidden layers and finally the output layer. This approach does not embrace functional data completely. \cite{wang2020nonlinear} recently extended FNN to model functional response, but again the architecture is limited, as it only embraces the functional data in the first hidden layer and the final output layer. As we can see above, much of the work are extensions of some version of scalar-on-function models. We propose a new architecture inspired from FNN and \cite{rao2021nonlinear} for modeling a functional response using functional inputs, where we have a continuous mapping throughout the network.

The major contribution of this work is solving the above problems by giving a novel framework that has no limitation with respect to the patterns it can learn while maintaining the functional nature of the data. Our framework consists of multiple continuous hidden layers consisting of continuous neurons that can learn functional representations from functional inputs to the functional outputs. We solve our framework by using two strategies in the function-on-function setting, {Function-on-Function} Direct Neural Networks (FFDNN) and {Function-on-Function} Basis Neural Networks (FFBNN). The rest of this paper is organized as follows: In Section 2, we first present some background on FDA and function-on-function models. Then we describe our framework and solve it using FFDNN and FFBNN. We end this section with penalization techniques to make our results much smoother and interpretable.  We follow this with simulations and several benchmark applications in Section 3. We consider multiple simulation settings depending on how the non-linear relation is defined between the functional response and functional predictors, and demonstrate the superiority of our methods over current state-of-the-art approaches. In the final section, we share concluding remarks and future research avenues which pertain to methodological and theoretical directions.

\section{Methodology}

We first give the necessary notation used in this paper. Let us assume that we have $N \in \mathcal{N}$ i.i.d subjects, where we observe functions \{($X_{i, 1}, \ldots, X_{i, R}),Y_i\}$, $i=1,\dots, N$, over a compact time interval $\mathcal{T}$. These functions
could be observed on other intervals, but as long as they are closed and bounded, they can always be rescaled to be on the unit interval [0,1]. For the $i^{th}$ subject, the functional inputs are $R$ random curves that can be denoted as $X_{i,r} \in \mathcal{L}^2(\mathcal{T})$ and the functional output is  $Y_{i}  \in \mathcal{L}^2(\mathcal{T})$ for $i=1, \ldots, N$ and $r=1, \ldots, R$.
We assume that all the functions are completely observed. This is known as dense functional data analysis \citep{FDA}. Our main objective is to learn the mapping, $F:\mathcal{L}^2(\mathcal{T}) \times \dots \times \mathcal{L}^2(\mathcal{T}) \to \mathcal{L}^2(\mathcal{T})$ from the functional inputs \{$X_{i,r}\}$ to the functional output $Y_{i}$:

\begin{equation}
\begin{split}
E[Y_{i}\mid X_{i,1},\ldots,X_{i, R} ]=F\left(X_{i, 1},\ldots,X_{i, R}\right)
\end{split}
\end{equation}

\subsection{Functional Regression}

We briefly describe the most common function-on-function regression models, which will be used later in simulations. The integrals without limits are defined over the entire domain. 

\begin{itemize}
    \item Function-on-function Linear Model (FFLM): 
    \begin{equation} \label{1}
    \begin{split}
    &\text{E}[Y_{i}(t)\mid X_{i, 1}, \ldots, X_{i, R} ]\\
    &=\alpha(t)+\sum_{r=1}^R\int \beta_r(s,t) X_{i,r}(s) ds, 
    \end{split}
    \end{equation}
    where $\alpha(t)$ is an intercept function and $\beta_r(s, t)$ are regression coefficient functions. 

    \item Continuously Additive Model (CAM): 
    \begin{equation} \label{2}
    \begin{split}
    &\text{E}[Y_{i}(t)\mid X_{i, 1}, \ldots, X_{i, R} ]\\
    &= \sum_{r=1}^R\int f(X_{i,r}(s),s,t)ds 
    \end{split}
    \end{equation}
    where the trivariate parameter function ${(x,s,t) }\rightarrow f(x,s,t)$ is smooth \citep{articlel4, reimherr2017optimal,Scheipl}. 

    \item Single Index Model:

    \begin{equation}\label{3}
    \begin{split}
    \text{E}[Y_{i}(t)\mid X_{i, 1}, \ldots, X_{i, R} ]=g\left(t,\sum_{r=1}^R \langle\beta_{r},X_{i,r}\rangle\right) 
    \end{split}
    \end{equation}

    where the function $g()$ is any {unknown} smooth function defined on the real line \citep{Jiang5}.
    
    \item Multiple Index Model:

    \begin{equation}\label{4}
    \begin{split}
    &\text{E}[Y_{i}(t)\mid X_{i, 1}, \ldots, X_{i, R} ]\\
    &=g\left(t, \sum_{r=1}^R \langle\beta_{1r},X_{i,r}\rangle,...,\sum_{r=1}^R\langle\beta_{Pr},X_{i,r}\rangle\right) 
    \end{split}
    \end{equation}
    
We define the above multiple Index models for simulation purposes, which is extension of the single index model {with $P$ indexes}. The scalar-on-function version of this model is well studied in \cite{articlem2,articlem3,article12,article13}.

\item Quadratic Model:
\begin{equation} \label{5}
\begin{split}
&\text{E}[Y_{i}(t)\mid X_{i, 1}, \ldots, X_{i, R} ] \\
& =\alpha(t)+\sum_{r=1}^R\int \beta_r(s,t) X_{i,r}(s) ds\\
&+\sum_{r=1}^R\int \int \beta_r(q,s,t) X_{i,r}(q)X_{i,r}(s) dqds, \\ 
\end{split}
\end{equation}

\begin{equation} \label{6}
\begin{split}
&\text{E}[Y_{i}(t)\mid X_{i, 1}, \ldots, X_{i, R} ]\\
&= \sum_{r=1}^R\int f(X_{i,r}(s),s,t)ds \\
& +\sum_{r=1}^R\int \int f(X_{i,r}(q),X_{i,r}(t),q,s,t)dqds 
\end{split}
\end{equation}

Equation \eqref{5} represents a Quadratic Model given by \cite{SUN2020106814} where $\alpha(t)$, $\beta(s,t)$ and $\beta(q,s,t)$ are unknown function coefficients. The Quadratic term is regarded as an interaction effect of the functional predictors on the functional response. For simulation purposes we define Equation \eqref{6} (Complex Quadratic Model).
    
\end{itemize}

These non-linear models are analogous to their multivariate and scalar-on-function counterparts, which help to overcome the curse of dimensionality {by using higher order terms or through indexes}. But these methods come with their own drawbacks like ignoring interactions, an inability to capture complex non-linear relationships, and restrictive structures. The true form of the relation between a functional response and the functional predictors in practice is difficult to identify. Hence, developing an approach that can deal with more general relationships between these functions without knowing the underlying ground truth is beneficial.

\subsection{Our Network Framework}

\begin{figure*}[]
\centering
\includegraphics[height=5.5cm]{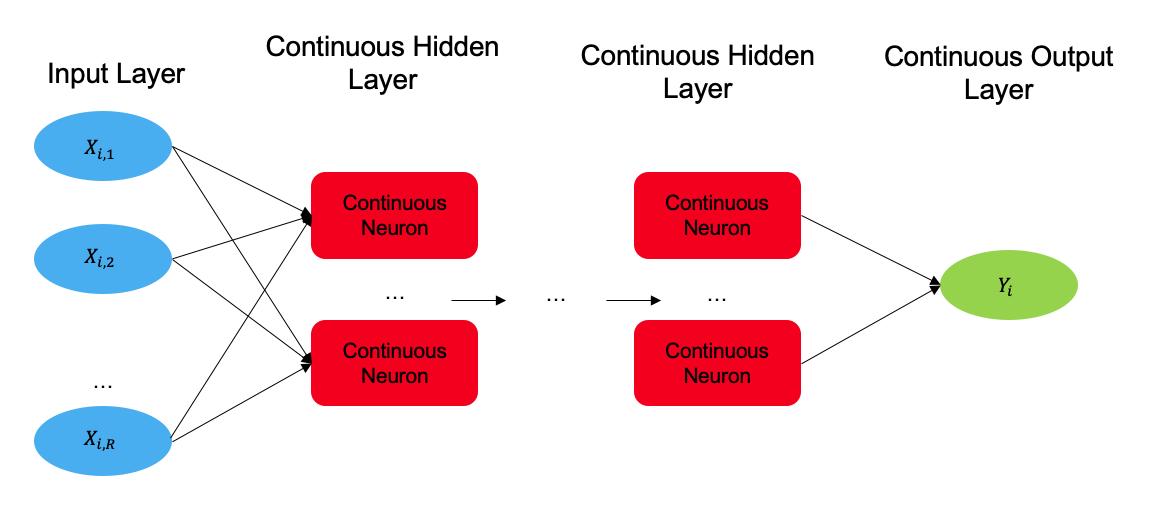}
  \caption{General architecture for our proposed approach with multiple continuous hidden layers.}
  \label{FFDNN2}
\end{figure*}

We propose a framework below, which develops a continuous mapping from layer to layer by using multiple \textit{continuous hidden layers} consisting of multiple \textit{continuous neurons}. In our framework, as seen in Figure \ref{FFDNN2}, we have three types of layers {that pass functions through the network}: an input layer, which is the first layer that takes in the functional predictors, the continuous hidden layers containing continuous neurons, and a continuous output layer that provides the final fitted functions using continuous neurons. We learn using the functional (univariate and bivariate) weights that are continuous over time (or other continuums). The main motivation for this approach is to preserve the functional nature of the data throughout the whole modeling process. This provides a richer structure to model functional data and also exploits the domain information and continuity of the predictor functions and response function. The $l^{th}$ continuous hidden layer and its $k^{th}$ continuous neuron is defined as:
\begin{equation}
\begin{split}
H^{(l)}_{(k)}(s)&=\sigma \Big(b^{(l)}_{(k)}(s) + \sum_{j=1}^{J} \int w^{(l)}_{(j,k)}(s,t)H^{(l-1)}_{(j)}(t)dt \Big) \label{e3}
\end{split}
\end{equation}
where $l=1,2,3,...,L$, $H^{(0)}_{(j)}(s)=X_j(s)$, $H^{(L)}(t)=\widehat{Y(t)}$, $b^{(l)}_{(k)} \in \mathcal{L}^2(\mcT)$ is the unknown intercept function, $w^{(l)}_{(j,k)} \in \mathcal L^2(\mcT \times \mcT)$ is the bivariate parameter function for the $k^{th}$ continuous neuron in the $l^{th}$ hidden layer coming from the $j^{th}$ continuous neuron of the $(l-1)^{th}$ hidden layer and $\sigma(\cdot)$ is a non-linear activation function. For simplicity and applicability of other methods, we have assumed a single output functions but our approach can easily accomodate for multiple response functions.

The proposed approach can be implemented either using FFDNN or FFBNN. FFDNN is a direct solution when the data is dense and is straightforward to implement. We learn through the functional weights and also have the flexibility to change the grid length (s) between the input and output layer. FFBNN uses basis expansions that allow the data to be irregular and learns through the basis weights, which allows for greater parsimony. 


\textbf{Functional Direct Neural Network (FFDNN)}: Let us assume that Figure \ref{FFDNN2} has $R$ inputs (functional predictors), $L$ continuous hidden layers, each with $J$ incoming and $K$ outgoing connections. FFDNN is similar to a neural network in nature for functional data, where we can define the number of continuous hidden layers and the number of continuous neurons in each continuous hidden layer. We use either ReLU, tanh, sigmoid/logistic, or linear (output layer) as the activation function in the continuous neurons. We reach from the functional predictors to the functional response via equation \eqref{e3}. While, using backpropagation (i.e. gradient decent), we learn the weight functions and the intercept functions.

We use the functional gradients given below to optimize for the network parameters. The gradient of the weight function measures the change in a functional to a change in a function on which the functional depends. \cite{1007599, wang2019multilayer,booko} give the necessary assumptions and mathematical tools (Fr\'echet derivatives) for getting the functional gradients. During the backpropagation step, we pass through the network backwards and we calculate the partial derivatives for the weight functions and the intercept functions. We define the loss function as $\mathcal{L}(\theta)$, 
where $\theta$  is the collection of all functional parameters. We define a quadratic loss function,
\begin{equation} \label{loss}
\begin{split}
&\mathcal{L}(\theta)=\frac{1}{N}  \sum_{i=1}^N \int \Big(Y_i(t)-\widehat{Y_i}(t)\Big)^2 dt.
\end{split}
\end{equation}
The partial derivatives needed for the parameters in our proposed architecture (as shown in Figure \ref{FFDNN2}) are as follows:
\begin{equation} \label{e5a}
\begin{split}
&\frac{\partial H^{(l)}_{(k)}}{\partial b^{(l)}_{(k)}}(s)\\
&= \sigma^\prime \Big(b^{(l)}_{(k)}(s) + \sum_{j=1}^{J} \int w^{(l)}_{(j,k)}(s,t)H^{(l-1)}_{(j)}(t)dt \Big)
\end{split}
\end{equation}

\begin{equation} \label{e5b}
\begin{split}
&\frac{\partial H^{(l)}_{(k)}}{\partial w^{(l)}_{(j,k)}}(s,t)\\
&=\sigma^\prime \Big(b^{(l)}_{(k)}(s) + \sum_{j^\prime=1}^{J} \int w^{(l)}_{(j^\prime,k)}(s,t^\prime)H^{(l-1)}_{(j^\prime)}(t^\prime)dt^\prime \Big)\\
&\times \int \frac{\partial}{\partial w^{(l)}_{(j,k)}(s,t)} w^{(l)}_{(j,k)}(s,t)H^{(l-1)}_{(j)}(t)dt \\
&=\sigma^\prime \Big(b^{(l)}_{(k)}(s) + \sum_{j^\prime=1}^{J} \int w^{(l)}_{(j^\prime,k)}(s,t^\prime)H^{(l-1)}_{(j^\prime)}(t^\prime)dt^\prime \Big)\\
&\times H^{(l-1)}_{(j)}(t)\\
\end{split}
\end{equation}
where  $l=1,2,3,...,L$, $H^{(0)}_{(j)}(s)=X_j(s)$, $H^{(L)}(t)=\widehat{Y}(t)$ and $\sigma^\prime(\cdot)$ represents the first derivative of $\sigma(\cdot)$.

In our architecture, the grid/domain ($s$), the number of continuous hidden layers, and the number of continuous neurons in each of the continuous hidden layer can be considered as hyperparameters. We optimize for the parameters in the same way for FFDNN as in classic neural networks, using backpropagation with the help of the gradients given above.

\textbf{Functional Basis Neural Network (FFBNN)}: In FFBNN, the continuous neurons are expanded using a basis expansion. The network learns the weights of the basis functions rather than the weight function itself. Basis functions can also help us to deal with irregular data more efficiently \citep{FDA}. In the forward propagation phase, FFBNN uses basis expansions where we replace the weight functions ($w^{(l)}_{(j,k)}(s,t)$) in Equation (\ref{e3}) by a linear combination of basis functions as shown below. Assume subscripts $j,k$ represent the continuous neurons of the neural network and superscript $l$ denote the $l^{th}$ layer for the weight parameter $(W^{(l)}_{(j,k)})$. The notation $v$ represents a standard collection of basis functions (splines, wavelets, or sine and cosine functions). Let us assume for simplicity, that we use the same basis functions and the same number of basis in each of the continuous neurons in the continuous hidden layers across the network. This gives us the following Equation:
\begin{equation} \label{e8}
\begin{split}
&H^{(l)}_{(k)}(s)=\sigma \Big(b^{(l)}_{(k)}(s) + \sum_{j=1}^{J} \int w^{(l)}_{(j,k)}(s,t)H^{(l-1)}_{(j)}(t)dt \Big)\\
&=\sigma \Big(\sum_{b=1}^{B} b^{(l)}_{(k),b} v^{*}_b(s) \\
&+ \sum_{j=1}^{J}  \sum_{c=1}^{C} \sum_{d=1}^{D}  w^{(l)}_{(j,k),c,d} v_c(s)  \int v_d(t) H^{(l-1)}_{(j)}(t)dt \Big)\\
&=\sigma \Big(\sum_{b=1}^{B} b^{(l)}_{(k),b} v^{*}_b(s) \\
&+ \sum_{j=1}^{J}  \sum_{c=1}^{C} \sum_{d=1}^{D}  {w^{(l)}_{(j,k),c,d}}  v_c(s)  A_{(j),d}^{(l)}\Big)\\
&=\sigma \Big(B^{(l)}_{(k)} v^{*}(s)  + \sum_{j=1}^{J}   W^{(l)}_{(j,k)}  v(s)  A_{(j)}^{(l)}\Big) \\
\end{split}
\end{equation}
where $l=1,2,3,...,L$, $H^{(0)}_{(j)}(s)=X_j(s)$ and $H^{(L)}(t)=\widehat{Y}(t)$. $B^{(l)}_{(k)}$, $W^{(l)}_{(j,k)}$ and $W^{(L+1)}_{(j)}$ are the unknown basis weights corresponding to their respective basis functions $v^{*}(s),$ $v(s),$ and $v(t)$. $B$, $C$ and $D$ specify the {number of basis functions} used in each of the continuous neuron of the continuous hidden layers. $\sigma(\cdot)$ is a non-linear activation function. 

Similar to FFDNN, the values of the grid/domain size ($s$), the number of continuous hidden layers, the number of continuous neurons in each continuous hidden layer, the selection of the basis functions, and the {number of basis functions} can be considered as hyperparameters. In FFBNN, we learn using the basis weights of the basis functions, which are required to reach from the functional predictors to the functional response via a chain of multiple CAMs in the forward propagation phase. The partial derivatives required for the parameters in our proposed architecture (as seen in Figure \ref{FFDNN2}) are as follows:
\begin{equation} \label{e10}
\begin{split}
&\frac{\partial H^{(l)}_{(j)}(s)}{\partial B^{(l)}_{(k)}}\\
&= \sigma^\prime \Big(B^{(l)}_{(k)}v^{*}(s) + \sum_{j=1}^{J}   W^{(l)}_{(j,k)}  v(s)  A_{(j)}^{(l)}\Big)\times v^{*}(s)\\
&\frac{\partial H^{(l)}_{(j)}(s)}{\partial W^{(l)}_{(j,k)}}\\
&= \sigma^\prime \Big(B^{(l)}_{(k)}v^{*}(s) + \sum_{j^\prime=1}^{J}   W^{(l)}_{(j^\prime,k)}  v(s)  A_{(j^\prime)}^{(l)}\Big) \times  v(s)  A_{(j)}^{(l)} 
\end{split}
\end{equation}
where l=1,2,3,...,L, $H^{(0)}_{(j)}(s)=X_j(s)$, $H^{(L)}(t)=\widehat{Y}(t)$, and $\sigma^\prime(\cdot)$ represents the first derivative of $\sigma(\cdot)$.

\subsection{Regularization}

We run into the problem of over-fitting with our approach like any other Machine Learning and Deep Learning method due to the overparameterization of the model. We first explore early stopping, which requires a validation set to stop the model fitting process early, before it starts to over-fit. Early stopping is straightforward to implement but splitting the training set further into validation set, can lead to a loss in performance, especially when the sample size is small. We instead consider a  cross-validated early stopping scheme to overcome this problem. We use k-fold cross-validation and note the early stopping iteration values in each fold along with the estimated parameter functions. We check the result of different early stopping schemes using central tendency of the iterations and the functional parameter values in the k-folds as shown in Appendix Table \ref{cv2} and Table \ref{cv3}.

Along with early stopping to deal with the over-fitting, we also consider a regularization approach by penalizing the roughness of the parameter functions in our framework. This regularization is called Roughness Penalization and is a more common regularization approach in FDA as well as other branches of nonparametric statistics. Interpreting black box models like our approach is a significant challenge in machine learning and deep learning, and can significantly reduce barriers to the adoption of the technology. We want to make use of the Roughness Penalty to move towards a more interpretable model. In this case, our loss function from Equation (\ref{loss}) changes to

\begin{equation} \label{lossreg}
\begin{split}
&\mathcal{L}_\lambda(\theta)=\frac{1}{N}  \sum_{i=1}^N \int \Big(Y_i(t)-\widehat{Y_i}(t)\Big)^2 dt\\
&+ \lambda_b   J_b(\theta_b)+ \lambda_w J_w(\theta_w)
\end{split}
\end{equation}
%
%
%
%
where $\lambda_b$ and $\lambda_w$ are the smoothing parameters.  The terms  $\mathcal J_b(\theta_b)$ and $\mathcal J_w(\theta_w)$ are used to penalize the intercept functions and bivariate functions respectively. We define $\mathcal J_b(\theta_b)$ and $\mathcal J_w(\theta_w)$  as the second order derivative and Laplacian operator of the parameter functions \citep{ramsay1997functional,FDA} as follows,
\begin{equation} \label{FFDNN_r}
\begin{split}
\mathcal J_b(\theta_b)&= \sum_{l=1}^L\sum_{k=1}^K\int \Big[{b^{(l)}_{(k)}}^{''}(s)\Big]^2ds\\
\mathcal J_w(\theta_w)&= \sum_{l=1}^L\sum_{k=1}^K\sum_{j=1}^J\int\int \Big[\Delta{w^{(l)}_{(j,k)}}(s,t)\Big]^2dsdt
\end{split}
\end{equation}
where
\begin{equation} \label{FFDNN_rl}
\Delta{w^{(l)}_{(j,k)}}(s,t)=\left(\frac{\partial^2}{\partial s^2} w_{j,k}^{(l)}(s,t) +\frac{\partial^2}{\partial t^2} w_{j,k}^{(l)}(s,t)\right)
\end{equation}

In the case of FFDNN, we are working on a grid, so we use numerical approximation to get the second order derivative of the parameter function and use zero-padding to match dimensions when necessary. For FFBNN, it is inherently smoother compared to FFDNN because of basis expansion, though the results can be rough depending on the selection of the basis functions, the number of basis functions used and the basis weight values. The penalty terms are the same in FFBNN but they are defined on the basis functions used to expand the parameter functions, i.e. we transfer all the derivatives and Laplacian operators to the basis functions \citep{kang2017manifold}. The regularization method is effective at improving the performance of FFDNN and FFBNN, as seen in the next section.

\section{Empirical Results}

In this section, we compare different approaches under multiple simulation setting to show the effectiveness of FFDNN and FFBNN, as well as bench marking with multiple real-world data sets. We demonstrate the following objectives through our results: 1) FFDNN and FFBNN are effectively trained by the derived optimization method 2) FFDNN and FFBNN enable learning of complex relation between the functional predictors and the functional response 3) FFDNN and FFBNN outperform other methods 4) Regularization leads to a smoother result.

\subsection{Simulations}

We consider a single predictor function (i.e. R=1) which is observed on a dense and regular grid. We generate $n=1100$ iid random curves $\left\{X_{1}(t), \cdots, X_{n}(t)\right\}$ from a Gaussian process with mean $0$ and covariance given as { $
  C_{X}(t, s)=\frac{\sigma^{2}}{\Gamma(\nu) 2^{\nu-1}}\left(\frac{\sqrt{2 \nu} \vert t-s \vert}{\rho}\right)^{\nu} K_{\nu}\left(\frac{\sqrt{2 \nu}\vert t-s \vert}{\rho}\right).
$} This is the Matérn covariance function and $K_{\nu}$ is the modified Bessel function of the second kind. We set $\rho=0.5,$ $\nu=5 / 2$ and $\sigma^{2}=1 .$ These curves are observed in an equally-spaced interval from $[0,1]$ at $m=100$ time points.

Table \ref{ss1} gives the different data generating models for the function-on-function setting. The response curves are evaluated at $m_y=75$ equally-spaced time points from $[0,1]$. Random noise  ($\epsilon_i(t)$) is added to our functional response model to adjust the signal to noise ratio. {Each simulation setting} is repeated 100 times and we divide the 1100 samples 
into 600 for training and 500 for testing. We use 100 samples from the training set as a validation set, which is required for Early Stopping in all the deep learning models (models that do not require early stopping use the 600 observations for training). In case of regularization using Laplace operator and basis weights, we use 5-fold cross-validation to tune the smoothing parameter $\lambda$. We report the Root Mean Square Error (RMSE) of the prediction using Equation (\ref{loss}). We apply traditional NN directly to the curves and FNN based on \cite{wang2019remaining}. For our approach, we consider multiple combinations of continuous hidden layers $(l=1, 2)$ and continuous neurons $(j=1, 2, 4)$. The grid $(s)$ value in the continuous neurons is in the range $(30, 300)$. The basis functions used in FFBNN are B-splines and the number of basis used is in the range $(5, 100)$.

We compare the performance of each method under different simulation settings with a functional response in Table \ref{t0}. We observe that only when the true model is Linear do all methods perform roughly the same, but as the relations become more complex and interactions are considered, many of the methods fail to capture the true relation between the functional predictor and the functional response. We see a substantial gain in such situations in our approach, which dominates all the other methods by giving the lowest RMSE (square root value of Equation \ref{loss}) values. The Complex Quadratic setting is especially interesting as the difference in performance in other methods to our method is substantial. The best possible RMSE is one (since that is the variance of the error function) and our method is performing well irrespective of the true nature between the functional predictor and the functional response.

\begin{table*}[]
\centering
\begin{tabular}{l} 
Simulation Scenarios \\
\hline 1. Linear:\text{ }$Y_{i}(t)=\alpha(t)+\int \beta(s,t) X_{i}(s) ds+\epsilon_i(t)$\\ 
\vspace{0.25cm}
$\text{where } \alpha=0, \beta(s,t)=\beta_1(s)*\beta_2(t),\text { } \beta_1(s)=5  \sin (2 \pi t),$
$\beta_2(t)=3  \sin ( 3 \pi t)$.\\

2. CAM:\text{ }$Y_i(t)=\int f(X_{i}(s),s,t)ds + \epsilon_i(t)$ \\
\vspace{0.25cm}
$\text {where } f(X_{i}(s),s,t)=X_{i}(s)^2st$\\

3. Single Index:$\text{ }Y_i(t)=g\left(t, \langle\beta,X_{i}\rangle\right) + \epsilon_i(t)$ \\
\vspace{0.25cm}
$\text{where } g(a,b)=a^0b^2, \text { }  \beta(s,t)=\beta_1(s)*\beta_2(t),\text { } \beta_1(s)=5  \sin (2 \pi t),$ $\beta_2(t)=3  \sin ( 3 \pi t)$\\ 

4. Multiple Index:$\text{ }Y_i(t)=g\left(t, \langle\beta_{1},X_{i}\rangle,\langle\beta_{2},X_{i}\rangle\right) + \epsilon_i(t) \quad$\\
$\text {where } g(a,b,c)=a^0b^2c^2, \text { }\beta_1(s,t)=\beta_a(s)*\beta_b(t),\text { } \beta_a(s)=5  \sin (2 \pi t),$ \\
\vspace{0.25cm}
$\beta_b(t)=3  \sin ( 3 \pi t), \text { } \beta_2(s,t) =\beta_c(s)*\beta_d(t), \text { } \beta_c(s)=4  \sin (5 \pi t),$ $\beta_d(t)=2  \sin ( 3 \pi t)$\\

5. Quadratic:$\text{ }Y_i(t)=\int \beta(s,t) X_{i}(s) ds+\int \int \beta(q,s,t) X_{i}(q)X_{i}(s) dqds + \epsilon_i(t) \qquad\qquad$\\
$\text {where } \alpha=0, \beta(s,t)=\beta_a(s)*\beta_b(t),\text { } \beta_a(s)=5  \sin (2 \pi s),\beta_b(t)=3  \sin (3 \pi s),$ \\
\vspace{0.25cm}
$\beta(q,s,t)=\beta_a(q)*\beta_b(s)*\beta_c(t), \text { } \beta_a(q)=5  \sin (3 \pi q)$ , $\beta_b(s)=5  \sin ( \pi s), \text { } \beta_c(t)=5  \sin ( \pi t)$\\

6. Complex Quadratic:$\text{ }Y_i(t)=\int f(X_{i}(s),s,t)ds +\int \int f(X_{i}(q),X_{i}(s),q,s,t)dqds + \epsilon_i(t) \qquad\qquad$\\
$\text  {where } f(X_{i}(s),s,t)=X_{i}(t)^2st,$\\ $f(X_{i}(q),X_{i}(s),q,s,t)=a^2$, a is the quadratic term from above.\\
\hline
\end{tabular}\\
\caption{Different simulation scenarios with functional response and a single predictor function.}
\label{ss1}
\end{table*}

\begin{table*}[]
\centering
\begin{tabular}{|l|l|l|l|l|l|l|l|}
\hline
Mapping & FFLM            & CAM            & NN & FNN  & FFDNN           & FFBNN                      \\ \hline
Linear          & {1.014} & {1.014} & 1.011    & {{1.007}}  & {{1.009}} & \textbf{{1.003}}               \\ \hline
CAM          & 1.732          & {1.409}  & 1.651 & 1.578    & {{1.225}} & \textbf{{1.213}}                   \\ \hline
Single Index           & 1.572           & 1.505          & 1.089    & 1.056 & \textbf{{1.019}} & {1.047}          \\ \hline
Multiple Index           & 2.106          & 1.875      & 2.123      & 1.854   & \textbf{{1.422}} & {1.552}   \\ \hline
Quadratic          & 1.672          & 1.221          & 1.072    & {{1.059}} & \textbf{{1.025}} & {{1.052}}      \\ \hline
Complex Quadratic          & 6.777          & 4.624          & 4.098    & 3.575 & \textbf{{2.293}} & {{2.633}}  \\ \hline
\end{tabular}
\caption{Comparison of the RMSE on the test set for the methods under different setting between the functional predictor and the functional response.}
\label{t0}
\end{table*}

\begin{figure*}[]
\centering
\includegraphics[height=5cm]{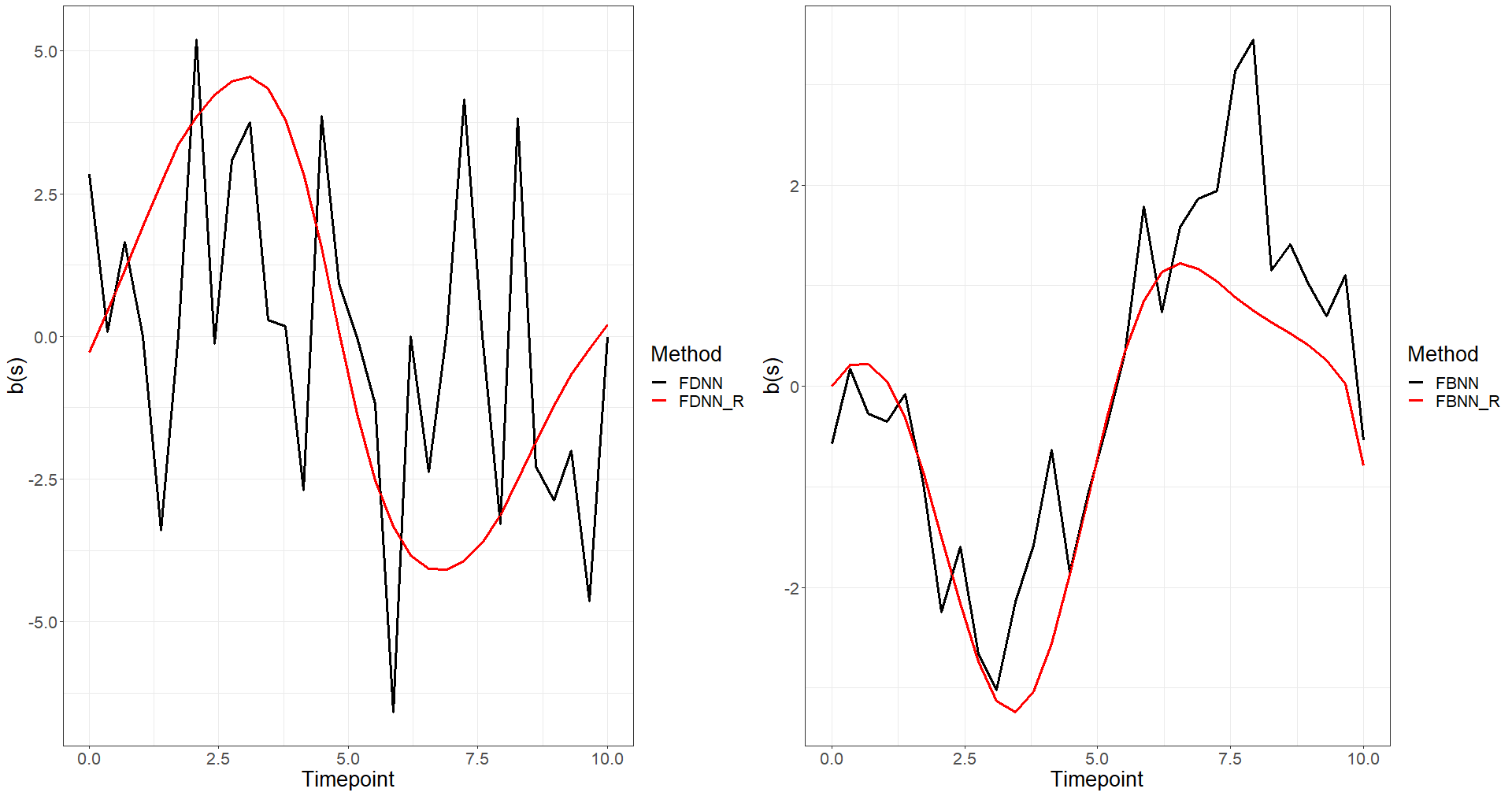}
  \caption{Smoothing for the intercept function $b(s)$ under regularization method.}
  \label{btr}
\end{figure*}

\begin{figure*}[]
\centering
\includegraphics[height=8cm]{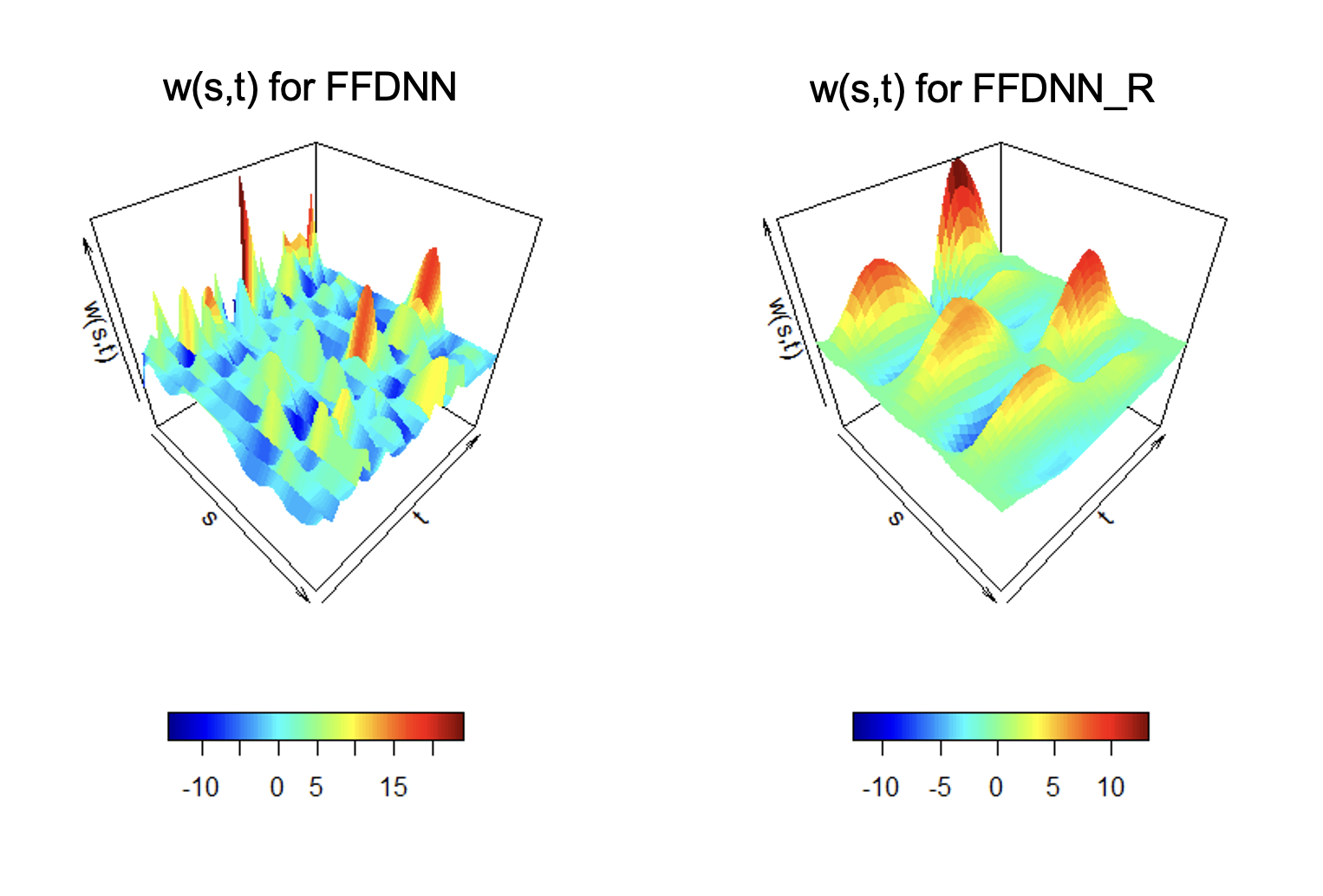}
  \caption{Smoothing for the bivariate function $w(s,t)$ using regularization method for FFDNN.}
  \label{wtrd}
\end{figure*}

\begin{figure*}[]
\centering
\includegraphics[height=8cm]{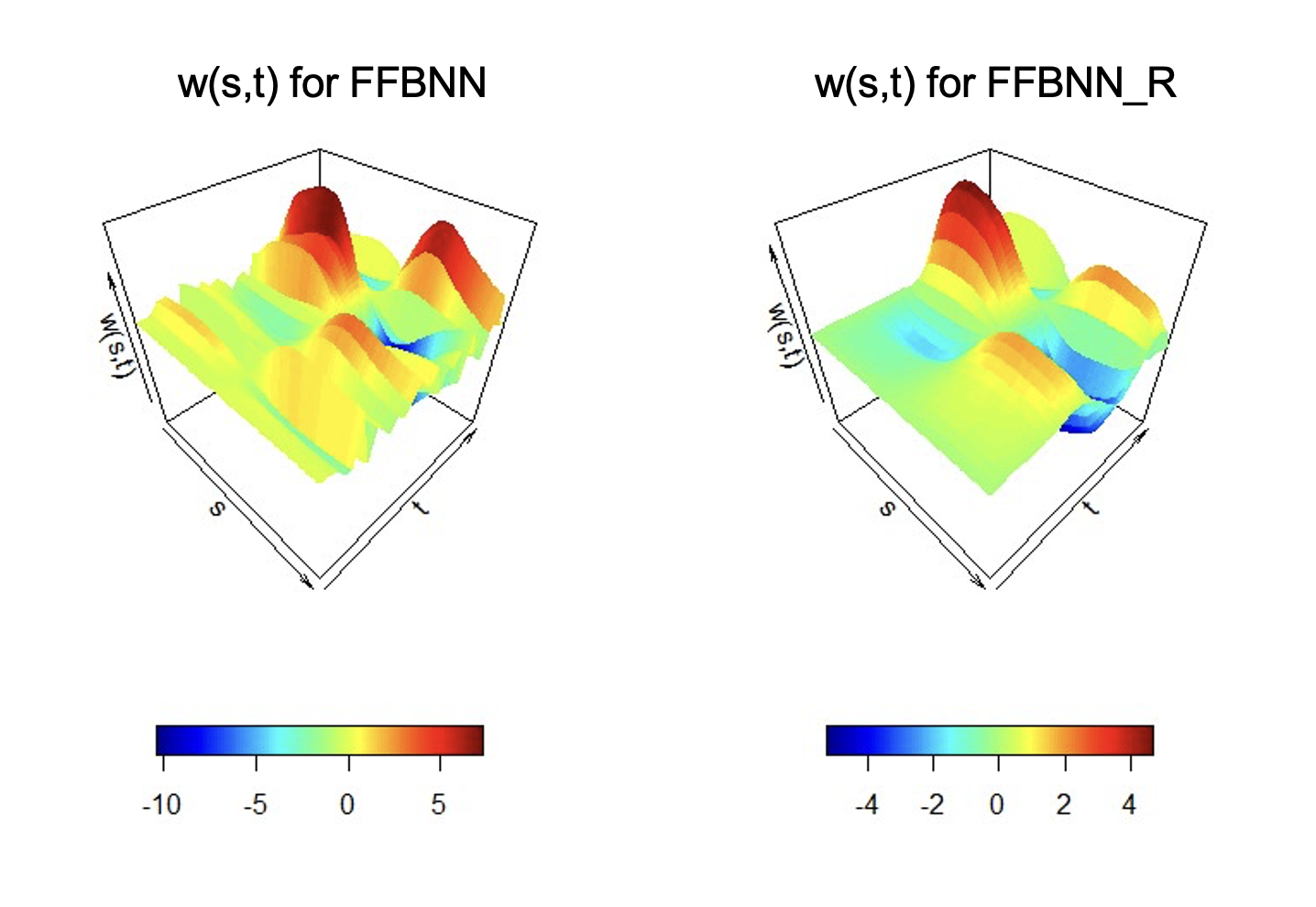}
  \caption{Smoothing for the bivariate function $w(s,t)$ under regularization method for FFBNN.}
  \label{wtrb}
\end{figure*}

Table \ref{r1} compares the result of our method under early stopping and regularization using the roughness penalty (denoted by FFDNN\_R and FFBNN\_R). We perform 5-fold cross-validation on the training set (600 samples) to tune the smoothing parameter ($\lambda$) in the regularization approach. The roughness penalty helps in preventing overfitting. We observe that the RMSE for the roughness penalty is slightly higher compared to the early stopping results but still competitive, and is better in comparison to other methods in Table \ref{t0}. This marginal decrease in performance for roughness penalty helps us to gain smoother results. Figure \ref{btr} represents the intercept function for an arbitrary continuous neuron in FFDNN and FFBNN with and without the roughness penalty. We can see that the roughness penalty intercept function captures the general shape very well while still maintaining reasonable performance in terms of RMSE. The bivariate function for FFDNN with and without the roughness penalty is shown in Figure \ref{wtrd}, where we can see a huge difference in the shape of the function, but the performance in terms of RMSE is similar. In the case of FFBNN, the bivariate function is relatively smoother compared to FFDNN without the roughness penalty and this is because we are using basis expansion. By adding the roughness penalty to FFBNN, we make the bivariate function much smoother and reduce it's range significantly as seen in Figure \ref{wtrb}. We again see that the bivariate function with roughness penalty captures the general shape very well in FFBNN and maintains reasonable performance in terms of RMSE as well. This is expected as we are penalizing the second derivative of the function. Regularization does not need a validation set, so there is no loss of information. Unlike early stopping, where we can stop at any intermediate iteration depending on the performance on the validation set, with roughness penalty, we can let the model stabilize to get the best result. The roughness penalty helps to get more reasonable parameter functions for each continuous neuron, opening up opportunities for interpretation and a better understanding of our approach. We check the performance for different early stopping strategies using 5-fold cross-validation but the results are similar with no clear winner as shown in Table \ref{cv2} and Table \ref{cv3}.(Appendix).

{We would also like to highlight the computational efficiency of various methods used, which is important for data analysis. FFLM proved to be the most time-efficient, with each iteration taking a mere 20 seconds, while CAM method, although slightly more time-consuming at 1.5 minutes per simulation, still maintained reasonable computational efficiency. The run times of the deep learning approaches vary significantly, contingent upon the architectural intricacies and other hyperparameters. We endeavored to maintain a consistent experimental environment, keeping the deep learning models as similar as possible. The standard Neural Network (NN) model required 45 seconds per simulation, whereas FNN needed 3 minutes, primarily due to the additional computational overhead incurred during the computation of Functional Principal Component Analysis (FPCA), which is integral to the FNN methodology. For our approaches, FFDNN needed 45 seconds for each iteration and the FFDBNN required 40 seconds per iteration. FFBNN runs slightly faster than FFDNN because the computational complexity is decreased with the use of basis expansion. Note that these experiments were conducted using R Software on a DELL XPS laptop equipped with an Intel Core i7 processor with 8 GB of memory.}

\begin{table*}[]
\centering
\begin{tabular}{|l|l|l|l|l|}
\hline
Mapping  & FFDNN & FFDNN\_R & FFBNN   & FFBNN\_R    \\ \hline
Linear  & 1.009 & 1.010 & 1.003     & 1.004              \\ \hline
CAM  & 1.225 & 1.328 & 1.213     & 1.335        \\ \hline
Single Index  & 1.019 & 1.043 & 1.047     & 1.116         \\ \hline
Multiple Index  & 1.422 & 1.549 & 1.552     & 1.631       \\ \hline
Quadratic  & 1.025 & 1.026 & 1.054     & 1.065         \\ \hline
Complex Quadratic  & 2.293 & 2.133 & 2.337     & 2.783         \\ \hline
\end{tabular}
\caption{{Comparison of the RMSE on the test set for our method with early stopping and reg-\\ularization under different setting between the functional predictor and the functional response.}}
\label{r1}
\end{table*}

\subsection{Real Data}

We analyze two different real data sets. The first example deals with the relation between electricity demand and temperature in Adelaide, Australia \citep{ade}, while the second examines how temperature for a day affects the bike rental patterns for that day \citep{bikedata}. {In both examples, the main motivation is to leverage weather patterns (which are readily available forecasts) to understand how temperature affects electricity demand and bike rentals}.

\begin{figure*}[]
\centering
\includegraphics[height=5cm]{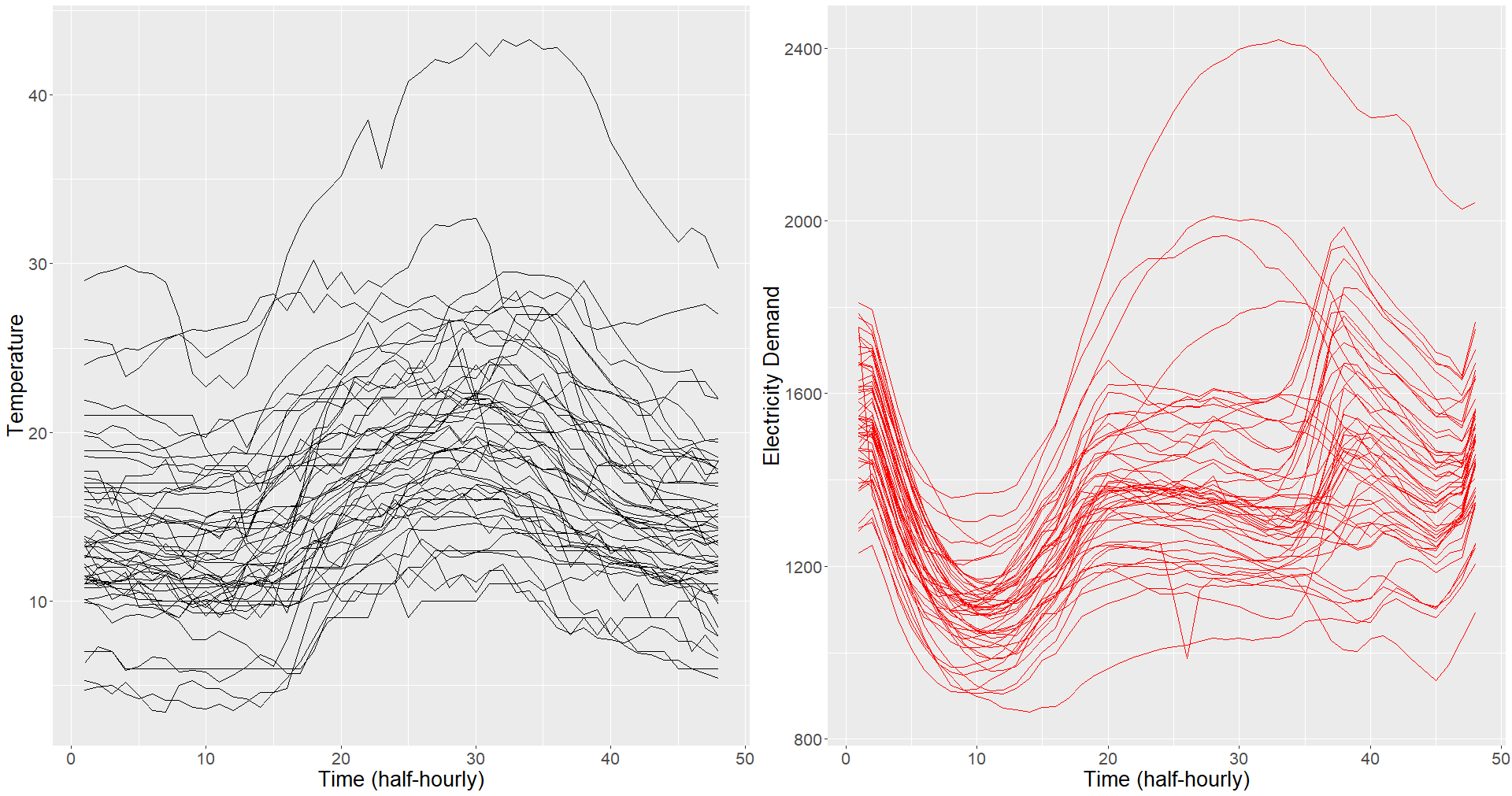}
  \caption{ Sample curves of temperatures in degrees (left panel) and the electricity demand (right panel) recorded half-hourly on Saturdays.}
  \label{rd1}
\end{figure*}

In the Adelaide data (available in $fds$ package in R), learning the variability of electricity demand in the city over time is of significance, as it can help to improve electricity storage, leading to cost reduction and sustainability. Several factors contribute to electricity demand but temperature can be considered as one of the most important variable related to it. Electricity demand has different dynamics on weekends compared to weekdays. We restrict our study to Saturday. Our focus is on how Saturday's electricity demand relates to the temperature. Therefore, we model the daily temperature to the daily electricity demand from 7/6/1997 to 3/31/2007 (508 days). Both temperature and electricity demand (Megawatts) are measured half-hourly, i.e. the functional predictor and the functional response are dense functional data measured at 48 time points in a day. Figure \ref{rd1} shows the half-hourly temperature and electricity demand for a sample of 50 days.

\begin{figure*}[]
\centering
\includegraphics[height=5cm]{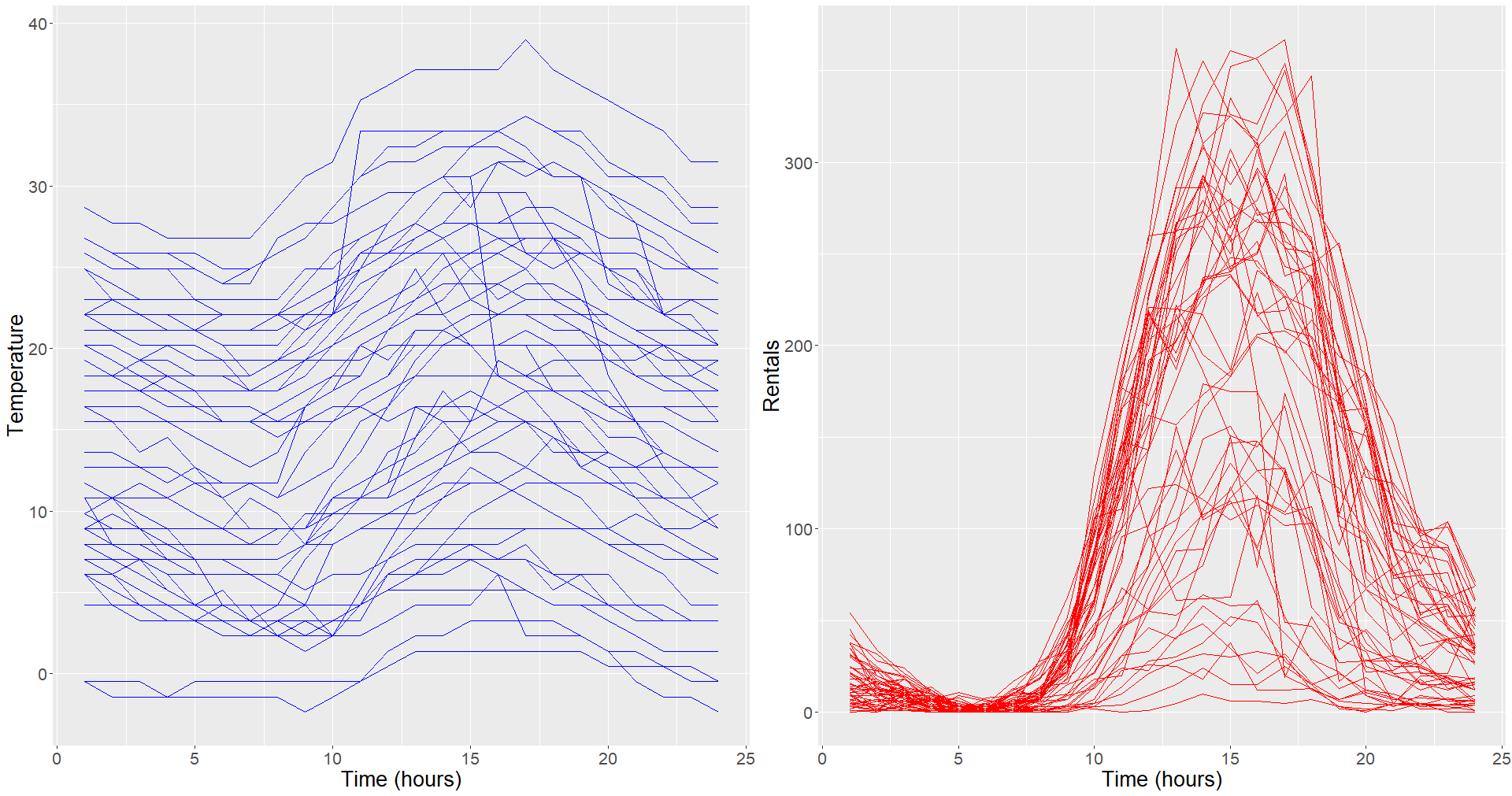}
  \caption{ Sample curves of temperatures in degrees (left panel) and the number of casual bike rentals (right panel) recorded hourly on Saturdays.}
  \label{rd2}
\end{figure*}

The bike data (available in the UCI data repository) consists of bike rental information for Washington, D.C., collected from the Capital Bike Share system. Bicycle rentals have become popular in recent years and are an attractive option to owing one. The goal is to learn the relation between bike demand and temperature so that the company avoids deploying too few or too many bikes than needed. We are interested in casual rentals per hour for a day, which are rentals to cyclists without membership. The bike rental dynamics vary on weekends compared to weekdays. We consider only Saturday rentals similar to the study in \cite{Kim1}. Hence, we model for the Saturday casual rentals using temperature information. We have the hourly rental and temperature information for the period of 01/01/2011 to 12/31/2012 (105 Saturdays). Since the measurements are hourly, we have 24 time points per day. Figure \ref{rd2} shows the counts of casual bike rentals and hourly temperature for 50 samples.

For the Adelaide data, we divide it into 408 training (80 validation) and 100 testing samples, and for the bike data, we have 85 training (16 validation) and 20 testing samples. Also, the bike data has a small amount of missing values, which requires smoothing the temperature curves using FPCA. We can see from Table \ref{tr1} that the true nature between the functional predictor and the functional response is non-linear and our methods perform the best when comparing the RMSE with other methods.

\begin{table*}[]
\centering
\begin{tabular}{|l|l|l|l|l|l|l|}
\hline
Data & FFLM   & CAM   & NN    & FNN   & FFDNN  & FFBNN  \\ \hline
Adelaide  & 188.376 & 140.917 & 208.296 & 180.816 & \textbf{{126.219}} & \textbf{{128.049}}\\ \hline
Bike   & 42.164 & 36.002 & 38.575 & 36.849 & \textbf{{33.141}} & \textbf{{33.113}}\\ \hline
\end{tabular}
\caption{Comparison of RMSE for different methods on the real data sets.}
\label{tr1}
\end{table*}

\section{Conclusions}

In this paper, we presented a novel approach for complex functional regression with a functional response. This approach was a natural extension of a neural network to functional data. We capture the temporal dependency and complicated relationships between the functional predictors and the functional response through the use of continuous hidden layers with continuous neurons. FFDNN and FFBNN don't assume any particular structure that limits the kind of non-linear patterns learned, especially, interactions within time points for the predictors. We also explored regularization techniques to get smoother results while maintaining the performance.

FFDNN and FFBNN are working not only on par with the best existing methods under simple cases but are outperforming in the case of non-linear settings. These approaches are appealing in practice as it provides a way for researchers to systematically conduct deep learning tasks when they have functional predictors and functional response. The main difference between FFDNN and FFBNN is that FFDNN is more straightforward to apply on dense and nicely gridded data with less tuning, whereas, FFBNN is more flexible and allows for greater parsimony with very dense grids as it only learns the basis weights. A deeper theoretical investigation of these approaches is needed and is an avenue for future work, to develop the statistical theory to give these methods a sound backing. Also, the performance of these approaches under a sparse setting can be of interest but requires nontrivial extensions.

Deep learning methods have been helping to make discoveries in multiple fields. In this paper, we merged the idea of neural network and functional data. There are multiple other methods in machine learning and deep learning that can be tailored for functional data like Clustering, Autoencoders, Convolutional Neural Network (CNN), Recurrent Neural Network (RNN), and many more. The applicability of such models is of huge significance as they can be used in many areas including Time Series, Natural Language Processing, and Computer Vision. Also, there is a lot of ongoing research for interpreting these black-box methods in deep learning. A major hurdle is defining interpretability for the problem in hand. While the roughness penalty is a step in the right direction, a deep dive into our approach for better understanding and interpretation can lead to more meaningful insights.

\section*{Acknowledgement}
This research was supported in part by  NSF SES-1853209.

\bibliographystyle{apalike}
\bibliography{arxiv}

\begin{thebibliography}{}

\bibitem[Cai and Yuan, 2012]{Caiff}
Cai, T.~T. and Yuan, M. (2012).
\newblock Minimax and adaptive prediction for functional linear regression.
\newblock {\em Journal of the American Statistical Association},
  107(499):1201--1216.

\bibitem[Chen et~al., 2012]{articlem2}
Chen, D., Hall, P., and Müller, H. (2012).
\newblock Single and multiple index functional regression models with
  nonparametric link.
\newblock {\em Annals of Statistics - ANN STATIST}, 39.

\bibitem[Conan-Guez and Rossi, 2002]{ConanGuez2002MultilayerPF}
Conan-Guez, B. and Rossi, F. (2002).
\newblock Multi-layer perceptrons for functional data analysis: A projection
  based approach.
\newblock In {\em ICANN}.

\bibitem[Eilers et~al., 2009]{EILERS2009196}
Eilers, P.~H., Li, B., and Marx, B.~D. (2009).
\newblock Multivariate calibration with single-index signal regression.
\newblock {\em Chemometrics and Intelligent Laboratory Systems}, 96(2):196 --
  202.
\newblock Chimiometrie 2007, Lyon, France, 29-30 November 2007.

\bibitem[Fan et~al., 2015]{articlel3}
Fan, Y., James, G.~M., and Radchenko, P. (2015).
\newblock Functional additive regression.
\newblock {\em Ann. Statist.}, 43(5):2296--2325.

\bibitem[Fanaee-T and Gama, 2013]{bikedata}
Fanaee-T, H. and Gama, J. (2013).
\newblock Event labeling combining ensemble detectors and background knowledge.
\newblock {\em Progress in Artificial Intelligence}, pages 1--15.

\bibitem[Ferraty, 2011]{bookr1}
Ferraty, F. (2011).
\newblock {\em Recent Advances in Functional Data Analysis and Related Topics}.

\bibitem[Ferraty et~al., 2013]{articlem3}
Ferraty, F., Goia, A., Salinelli, E., and Vieu, P. (2013).
\newblock Functional projection pursuit regression.
\newblock {\em TEST}, 22.

\bibitem[Ferraty and Vieu, 2006]{10}
Ferraty, F. and Vieu, P. (2006).
\newblock {\em Nonparametric Functional Data Analysis: Theory and Practice
  (Springer Series in Statistics)}.
\newblock Springer-Verlag, Berlin, Heidelberg.

\bibitem[Horv{\'a}th and Kokoszka, 2012]{hor}
Horv{\'a}th, L. and Kokoszka, P. (2012).
\newblock {\em Inference for Functional Data with Applications}.
\newblock Springer Series in Statistics. Springer New York.

\bibitem[James and Silverman, 2004]{articlem1}
James, G. and Silverman, B. (2004).
\newblock Functional adaptive model estimation.
\newblock {\em Journal of the American Statistical Association}, 100.

\bibitem[Jiang and Wang, 2011]{Jiang5}
Jiang, C.-R. and Wang, J.-L. (2011).
\newblock {Functional single index models for longitudinal data}.
\newblock {\em The Annals of Statistics}, 39(1):362 -- 388.

\bibitem[Kadri et~al., 2010]{Kadri}
Kadri, H., Duflos, E., Preux, P., Canu, S., and Davy, M. (2010).
\newblock Nonlinear functional regression: a functional rkhs approach.
\newblock In Teh, Y.~W. and Titterington, M., editors, {\em Proceedings of the
  Thirteenth International Conference on Artificial Intelligence and
  Statistics}, volume~9 of {\em Proceedings of Machine Learning Research},
  pages 374--380, Chia Laguna Resort, Sardinia, Italy. PMLR.

\bibitem[Kang et~al., 2017]{kang2017manifold}
Kang, H.~B., Reimherr, M., Shriver, M., and Claes, P. (2017).
\newblock Manifold data analysis with applications to high-frequency 3d
  imaging.

\bibitem[Kim et~al., 2018]{Kim1}
Kim, J.~S., Staicu, A.-M., Maity, A., Carroll, R.~J., and Ruppert, D. (2018).
\newblock Additive function-on-function regression.
\newblock {\em Journal of Computational and Graphical Statistics},
  27(1):234--244.
\newblock PMID: 29780218.

\bibitem[Kokoszka and Reimherr, 2018]{FDA}
Kokoszka, P. and Reimherr, M. (2018).
\newblock {\em Introduction to Functional Data Analysis}.
\newblock New York: Chapman and Hall/CRC.

\bibitem[Lian, 2007]{Lian}
Lian, H. (2007).
\newblock Nonlinear functional models for functional responses in reproducing
  kernel hilbert spaces.
\newblock {\em Canadian Journal of Statistics}, 35(4):597--606.

\bibitem[Ma and Zhu, 2016]{articlel4}
Ma, H. and Zhu, Z. (2016).
\newblock Continuously dynamic additive models for functional data.
\newblock {\em Journal of Multivariate Analysis}, 150:1 -- 13.

\bibitem[Magnano et~al., 2008]{ade}
Magnano, L., Boland, J.~W., and Hyndman, R.~J. (2008).
\newblock Generation of synthetic sequences of half-hourly temperature.
\newblock {\em Environmetrics}, 19(8):818--835.

\bibitem[McLean et~al., 2014]{McLean}
McLean, M.~W., Hooker, G., Staicu, A., Scheipl, F., and Ruppert, D. (2014).
\newblock Functional generalized additive models.
\newblock {\em Journal of Computational and Graphical Statistics},
  23(1):249--269.

\bibitem[Morris, 2014]{articler1m}
Morris, J. (2014).
\newblock Functional regression.
\newblock {\em Annual Review of Statistics and Its Application}, 2.

\bibitem[Muller et~al., 2013]{articlel1}
Muller, H., Wu, Y., and Yao, F. (2013).
\newblock {Continuously additive models for nonlinear functional regression}.
\newblock {\em Biometrika}, 100(3):607--622.

\bibitem[Muller and Yao, 2008]{article12}
Muller, H. and Yao, F. (2008).
\newblock Functional additive models.
\newblock {\em Journal of the American Statistical Association},
  103:1534--1544.

\bibitem[Olver, 2019]{booko}
Olver, P.~J. (2019).
\newblock {\em Introduction to the calculus of variations}.

\bibitem[Preda, 2007]{PREDA2007829}
Preda, C. (2007).
\newblock Regression models for functional data by reproducing kernel hilbert
  spaces methods.
\newblock {\em Journal of Statistical Planning and Inference}, 137(3):829--840.
\newblock Special Issue on Nonparametric Statistics and Related Topics: In
  honor of M.L. Puri.

\bibitem[Ramsay and Silverman, 1997]{ramsay1997functional}
Ramsay, J. and Silverman, B. (1997).
\newblock {\em Functional Data Analysis}.
\newblock Springer series in statistics. Springer.

\bibitem[Rao and Reimherr, 2021]{rao2021nonlinear}
Rao, A.~R. and Reimherr, M. (2021).
\newblock Non-linear functional modeling using neural networks.

\bibitem[Reimherr et~al., 2017]{reimherr2017optimal}
Reimherr, M., Sriperumbudur, B., and Taoufik, B. (2017).
\newblock Optimal prediction for additive function-on-function regression.

\bibitem[Reiss et~al., 2016]{articler1}
Reiss, P., Goldsmith, J., Shang, H.~L., and Ogden, R. (2016).
\newblock Methods for scalar‐on‐function regression.
\newblock {\em International Statistical Review}, 85.

\bibitem[Rossi and Conan-Guez, 2006]{articler4}
Rossi, F. and Conan-Guez, B. (2006).
\newblock Theoretical properties of projection based multilayer perceptrons
  with functional inputs.
\newblock {\em Neural Processing Letters}, 23.

\bibitem[{Rossi} et~al., 2002]{1007599}
{Rossi}, F., {Conan-Guez}, B., and {Fleuret}, F. (2002).
\newblock Functional data analysis with multi layer perceptrons.
\newblock In {\em Proceedings of the 2002 International Joint Conference on
  Neural Networks. IJCNN'02 (Cat. No.02CH37290)}, volume~3, pages 2843--2848
  vol.3.

\bibitem[Scheipl et~al., 2015]{Scheipl}
Scheipl, F., Staicu, A.-M., and Greven, S. (2015).
\newblock Functional additive mixed models.
\newblock {\em Journal of Computational and Graphical Statistics},
  24(2):477--501.

\bibitem[Stoker, 1986]{s1}
Stoker, T.~M. (1986).
\newblock Consistent estimation of scaled coefficients.
\newblock {\em Econometrica}, 54(6):1461--1481.

\bibitem[Sun et~al., 2018]{Sun}
Sun, X., Du, P., Wang, X., and Ma, P. (2018).
\newblock Optimal penalized function-on-function regression under a reproducing
  kernel hilbert space framework.
\newblock {\em Journal of the American Statistical Association},
  113(524):1601--1611.
\newblock PMID: 30799886.

\bibitem[Sun and Wang, 2020]{SUN2020106814}
Sun, Y. and Wang, Q. (2020).
\newblock Function-on-function quadratic regression models.
\newblock {\em Computational Statistics \& Data Analysis}, 142:106814.

\bibitem[Wang et~al., 2020]{wang2020nonlinear}
Wang, Q., Wang, H., Gupta, C., Rao, A.~R., and Khorasgani, H. (2020).
\newblock A non-linear function-on-function model for regression with time
  series data.

\bibitem[Wang et~al., 2019]{wang2019remaining}
Wang, Q., Zheng, S., Farahat, A., Serita, S., and Gupta, C. (2019).
\newblock Remaining useful life estimation using functional data analysis.

\bibitem[{Wang} et~al., 2019]{wang2019multilayer}
{Wang}, Q., {Zheng}, S., {Farahat}, A., {Serita}, S., {Saeki}, T., and {Gupta},
  C. (2019).
\newblock Multilayer perceptron for sparse functional data.
\newblock In {\em 2019 International Joint Conference on Neural Networks
  (IJCNN)}, pages 1--10.

\bibitem[Wang and Ruppert, 2013]{articlel2}
Wang, X. and Ruppert, D. (2013).
\newblock Optimal prediction in an additive functional model.
\newblock {\em Statistica Sinica}, 25.

\bibitem[Yao and Muller, 2010]{quad}
Yao, F. and Muller, H. (2010).
\newblock Functional quadratic regression.
\newblock {\em Biometrika}, 97(1):49--64.

\bibitem[Zhu et~al., 2013]{article13}
Zhu, H., Yao, F., and Zhang, H. (2013).
\newblock Structured functional additive regression in reproducing kernel
  hilbert spaces.
\newblock {\em Journal of the Royal Statistical Society: Series B (Statistical
  Methodology)}, 76.

\end{thebibliography}

\newpage
\section{Appendix}

In Table \ref{cv2} and Table \ref{cv3}, we compare the results of different early stopping strategies using 5-fold Cross-Validation (CV) without a validation set against the results from early stopping with validation set. The strategies we use are, taking the mean, median, maximum and minimum of the early stopping iteration values between the 5-folds and also the average of the functional parameter values across the 5-folds denoted by Wavg. As seen in the tables below, the results are inconsistent and there is no clear winner. Also, interestingly the difference in performance is more for FFBNN compared to FFDNN. For now, the mean and median strategy looks to be the most promising among the other options. A more holistic investigation is required to find the best option. One reason might be the black-box nature of our approach, as depending on the architecture of the network, step size, and initialization of the functions, different folds are learning at a different rate and are stopping early or much later. This could be causing under-fitting in the overall model.

\begin{table*}[h]
\centering
\begin{tabular}{|l|l|l|l|l|l|l|}
\hline
\begin{tabular}[c]{@{}l@{}}Mapping\\   \end{tabular} & \begin{tabular}[c]{@{}l@{}}FFDNN\\   Mean\end{tabular} & \begin{tabular}[c]{@{}l@{}}FFDNN\\   Med\end{tabular} & \begin{tabular}[c]{@{}l@{}}FFDNN\\   max\end{tabular} & \begin{tabular}[c]{@{}l@{}}FFDNN\\   min\end{tabular} & \begin{tabular}[c]{@{}l@{}}FFDNN\\   Wavg\end{tabular} & \begin{tabular}[c]{@{}l@{}}FFDNN\\   \end{tabular}      \\ \hline
Linear                                             & 3.054                                                 & 3.054                                                & 1.949                                                & 3.054                                                & 1.034                                                 & 1.009                                                          \\ \hline
CAM                                             & 1.083                                                 & 1.099                                                & 1.089                                                & 1.087                                                & 1.114                                                 & 1.225                                              \\ \hline
Single Index                                            & 1.026                                                 & 1.027                                                & 1.043                                                & 1.019                                                & 1.168                                                 & 1.028                                                      \\ \hline
Multiple Index                                             & 1.550                                                 & 1.563                                                & 1.573                                                & 1.692                                                & 2.164                                                 & 1.422                                                      \\ \hline
Quadratic                                            & 1.094                                                 & 1.175                                                & 1.093                                                & 1.202                                                & 1.420                                                 & 1.025                                                        \\ \hline
Complex Quadratic                                             & 3.037                                                 & 3.030                                                & 3.017                                                & 2.293                                                & 3.010                                                  & 2.316                                                                                                                    \\ \hline

\end{tabular}
\caption{Comparison of the RMSE for FFDNN method with multiple early stopping strategies using CV under different setting between the functional predictor and the functional response.}
\label{cv2}
\end{table*}

\begin{table*}[h]
\centering
\begin{tabular}{|l|l|l|l|l|l|l|l|l|l|l|l|l|}
\hline
\begin{tabular}[c]{@{}l@{}}Mapping\\   \end{tabular}  & \begin{tabular}[c]{@{}l@{}}FFBNN\\   Mean\end{tabular} & \begin{tabular}[c]{@{}l@{}}FFBNN\\   Med\end{tabular} & \begin{tabular}[c]{@{}l@{}}FFBNN\\   Max\end{tabular} & \begin{tabular}[c]{@{}l@{}}FFBNN\\   Min\end{tabular} & \begin{tabular}[c]{@{}l@{}}FFBNN\\   Wavg\end{tabular} & \begin{tabular}[c]{@{}l@{}}FFBNN\\   \end{tabular}     \\ \hline
Linear                                                                                         & 1.324                                                 & 1.352                                                & 1.197                                                & 1.208                                                & 1.625                                                 & 1.003                                           \\ \hline
CAM                                                                                       & 1.060                                                 & 1.060                                                & 1.059                                                & 1.064                                                & 1.356                                                 & 1.213                                                  \\ \hline
Single Index                                                                                   & 1.378                                                 & 1.378                                                & 1.378                                                & 1.378                                                & 1.381                                                 & 1.047                                         \\ \hline
Multiple Index                                                                                     & 2.069                                                 & 2.085                                                & 2.043                                                & 2.426                                                & 2.467                                                 & 1.552                                            \\ \hline
Quadratic                                                                                       & 1.717                                                 & 1.717                                                & 1.717                                                & 1.633                                                & 1.608                                                 & 1.052                                               \\ \hline
Complex Quadratic                                                                                       & 3.045                                                 & 3.002                                                & 2.034                                                & 2.962                                                & 2.958                                                 & 2.633                                            \\ \hline

\end{tabular}
\caption{Comparison of the RMSE for FFBNN method with multiple early stopping strategies using CV under different setting between the functional predictor and the functional response.}
\label{cv3}
\end{table*}

\end{document}